\documentclass[a4paper, nofootinbib, prd, preprintnumbers, showpacs, twocolumn]{revtex4}

\usepackage{amsfonts}
\usepackage{amssymb}
\usepackage[english]{babel}
\usepackage{exscale}
\usepackage[T1]{fontenc}
\usepackage[final]{graphicx}
\usepackage[latin9]{inputenc}
\usepackage{mathpple}
\usepackage{textcomp}

\graphicspath{{plots/headon/}{plots/meudon/}{plots/pi0/}}

\newcommand{\mtheta}{\ensuremath{\Theta}}
\newcommand{\mthetar}{\ensuremath{r \Theta}}
\newcommand{\mthetarr}{\ensuremath{r^2 \Theta}}
\newcommand{\mthetarbar}{\ensuremath{r}-bar \ensuremath{\Theta}}
\newcommand{\msmallestr}{\emph{smallest} \ensuremath{r}}
\newcommand{\minnerthetar}{\emph{inner} \ensuremath{r \Theta}}

\begin{document}

\preprint{AEI-2004-089}

\title{Horizon Pretracking}

\author{Erik Schnetter}
\email{schnetter@aei.mpg.de}

\author{Frank Herrmann}
\email{herrmann@aei.mpg.de}

\author{Denis Pollney}
\email{pollney@aei.mpg.de}

\affiliation{Max-Planck-Institut für Gravitationsphysik,
  Albert-Einstein-Institut, Am Mühlenberg 1, D-14476 Potsdam, Germany}
\homepage{http://numrel.aei.mpg.de/}

\date{February 1, 2005}


\begin{abstract}
  We introduce \emph{horizon pretracking} as a method for analysing
  numerically generated spacetimes of merging black holes.
  Pretracking consists of following certain modified constant
  expansion surfaces
  during a simulation before a common apparent horizon has
  formed.  The tracked surfaces exist at all times, and are
  defined so as to include the common apparent horizon if it exists.
  The method provides a way for finding this common apparent
  horizon in an efficient and reliable manner at the earliest possible
  time.  
  We can distinguish inner and outer horizons by examining the
  distortion of the surface.
  Properties of the pretracking surface such as its
  expansion, location, shape, area, and angular momentum can also be
  used to predict when a common apparent horizon will appear, and
  its characteristics.  The latter could also be used to
  feed back into the simulation by adapting e.g.\ boundary or gauge
  conditions even before the common apparent horizon has formed.
\end{abstract}

\maketitle

\section{Motivation}

In a spacetime that contains coalescing binary black holes,
and given suitable gauge conditions, a common
apparent horizon forms some time after the event horizons have merged.
Similarly, in a spacetime containing a collapsing star, an apparent
horizon forms some time after the event horizon has formed.
In numerical calculations it is important to locate apparent
horizons.  On one hand, one can extract interesting physical
information from an apparent horizon, such as the object's mass and
spin, for instance via the \emph{dynamical horizon}
formalism~\cite{Ashtekar:2004cn, Dreyer-etal-2002-isolated-horizons}.
On the other
hand, certain numerical methods such as excision boundaries
\cite{Cook97a, Alcubierre01a} use the apparent horizon to identify the
region within the black hole. Knowing the common horizon location also
makes it possible to adjust the dynamics of the spacetime via
so-called ``horizon-locking'' gauges~\cite{Anninos94e,
Balakrishna96a}, as well as gauges which attempt to control the
location of the black hole~\cite{Alcubierre2003:co-rotating-shift}.
At late times, the shape and
oscillations of the apparent horizon is an effective indicator of the
angular momentum of the final black hole, and has been related to its
quasi-normal mode ringing~\cite{Anninos94b}.

From a practical standpoint, it is the apparent horizons which are
more interesting than the event horizon at a given instant of time,
since the latter is a globally defined quantity whose location can
only be determined once the entire future development of the spacetime
is known. The apparent horizon, on the other hand, is defined by the
outermost marginally trapped surface and can be found locally in time.
Further,
it is always contained within the event horizon~\cite{Hawking73a}
when the strong energy condition holds,
making it a ``safe'' estimator of the black hole location
for use by excision boundaries~\cite{Thornburg87, Seidel92a}.

A typical simulation scenario involves a pair of binary black holes,
identified by disconnected horizons in the initial data surface,
evolving until they come close enough that a \emph{common horizon}
forms around them.  The common apparent horizon appears
instantaneously as a surface enveloping the two individual
horizons. This is different from the situation for event horizons,
where the disconnected lobes ``grow together'' to form a single
connected surface~\cite{Hawking73a, Anninos94b}.  For a given
dynamical evolution, it is not known
\emph{a priori} when and where a common apparent horizon first
appears, as this depends on both the initial data as well as the particular
slicing of the spacetime.

Fast apparent horizon finders, which essentially solve a nonlinear
elliptic equation, require a good initial guess for both the location
and shape of the surface \cite{Thornburg95}.  For horizons which are
present in the spacetime from the initial time, it is usually
sufficient to use the previously detected surface as an estimate for
the horizon location at some small time interval later. The initial data
construction itself often provides a good first guess for
individual horizons.  The common horizon, however, appears
instantaneously at some late time and without a previous good guess
for its location.  In practice, an estimate of the surface location
and shape can be put in by hand. The quality of this guess will
determine the rate of convergence of the finder and, more seriously,
also determines whether a horizon is found at all. Gauge effects in the
strong field region can induce distortions that have a large influence
on the shape of the common horizon, making them difficult to predict,
particularly after a long evolution using dynamical coordinate
conditions. As such, it can be a matter of some expensive trial and
error to find the common apparent horizon at the
earliest possible time. Further, if a common apparent horizon is not
found, it is not clear whether this is because there is none, or
whether there exists one which has only been missed due to unsuitable
initial guesses --- for a fast apparent horizon finder, a good initial
guess is crucial.

In this paper, we present a reliable method for determining the first
common horizon to appear in a dynamical spacetime evolution.
The method provides a close to optimal initial guess for the horizon surface
and reliable predictions of its physical properties.
Roughly speaking,
\emph{horizon pretracking}~\cite{Schnetter03a}
involves searching the spacetime for the
surface of minimal constant expansion which envelops the source.  For
a spacetime without a common horizon, this surface will have positive
expansion.  Assuming that a common horizon does form after some time,
the expansion of the pretracking surface will gradually decrease to
zero with time, at which point the common horizon has formed, by
definition.

The rationale for this method is that it is much more reliable and
efficient to ``track'' a surface than to find an entirely new
surface, the difference being that a good initial guess is provided by
the previous value of the tracked surface.  In the past, a method such
as pretracking might have been too computationally expensive to be
implemented as a practical solution. However, careful consideration of
the problem of finding constant expansion surfaces has resulted in
algorithms which typically solve the system in a matter of seconds
\cite{Thornburg95, Schnetter03a, Thornburg2003:AH-finding}.  Thus the
required search for the minimal expansion surface becomes feasible.

Pretracking has some similarities to minimisation \cite{Anninos98b,
Metzger04} or
curvature \cite{Shoemaker-Huq-Matzner-2000} flow methods for apparent
horizon finding.  These methods solve a minimisation problem or a
parabolic equation to find horizons, starting from a large sphere, and
can give a reliable answer as to whether a common horizon exists.
Their disadvantage is that these methods are extremely slow and cannot
be applied at every time step.  Pretracking is faster since it
``tracks'' instead of minimising or ``flowing'' at each time.
Additionally, the pretracking surfaces can provide valuable
information about the current state of the simulation.

In Section~\ref{sec:background} we briefly discuss surfaces of
constant expansion.  Section~\ref{sec:pretracking} defines the notion
of a pretracking surface, and outlines a number of algorithmic details
involving the parametrisation of the surface, and the binary search
method for the minimal constant expansion surfaces.  Finally in
Section~\ref{sec:results} we present applications of the technique to
binary black hole mergers.

\section{Apparent Horizons and Constant Expansion Surfaces}
\label{sec:background}

An apparent horizon (AH) is the outermost marginally outer-trapped
surface in a spacelike hypersurface $\Sigma$.  The AH satisfies the
conditions $\Theta_{(\ell)} = 0$ and $\Theta_{(n)}<0$, where
$\Theta_{(\ell)}$ and $\Theta_{(n)}$ are the expansions of the
outgoing and ingoing null normals of the surface.  These can be
calculated from the ADM variables via
\begin{eqnarray}
  \Theta_{(\ell)} & = & + \nabla_i s^i + K_{ij} (s^i s^j -
  \gamma^{ij}) \quad\textrm{and}
  \\
  \Theta_{(n)} & = & - \nabla_i s^i + K_{ij} (s^i s^j - \gamma^{ij})
  \textrm{.}
\end{eqnarray}
Here $\gamma_{ij}$ and $K_{ij}$ are the three-metric and extrinsic
curvature of the spacelike hypersurface, and $s_i$ is the spacelike
normal to the surface within the spacelike hypersurface.

We describe the surface explicitly through a function $h(\theta,\phi)$
which specifies the radius of the surface as function of the angular
coordinates $\theta$ and $\phi$ about some origin.  The spacelike
normal is then given by $s_i = \nabla_i F / | \nabla F |$, where
$F(r,\theta,\phi) = r - h(\theta,\phi)$ with the coordinate radius
$r$, hence $F(r,\theta,\phi)=0$ indicates the surface location.  This
explicit representation is convenient but not necessary for surfaces
with $S^2$ topology, and is not restrictive in practice
\cite{Thornburg2003:AH-finding}.

The notion of pretracking requires specifying a general family of
surfaces for spacetimes which may not have an apparent horizon, but
which contain the AH as a surface if one exists. For instance, a
generalisation to the concept of apparent horizons are the
\emph{constant expansion (CE)} surfaces, which are defined by the
condition
\begin{eqnarray}
  \Theta_{(\ell)} & = & C,
\end{eqnarray}
with $C$ a constant over the surface~\cite{Schnetter03a}.
The apparent horizon is a CE surface with $C=0$.  For $K_{ij}=0$, the CE
surfaces are identical to \emph{constant mean curvature (CMC)}
surfaces, which are defined by $\nabla_i s^i = C$.  In an
asymptotically flat spacetime with zero extrinsic curvature, the CMC
surfaces foliate the hypersurface in a neighbourhood of
infinity~\cite{Huisken96}.
Thus a parametrised family of CMC surfaces can be specified at large
radii, though they may not exist in the strong field interior of a
spacelike slice.  In numerical experiments with
isolated sources, we have found that CE surfaces form a similarly
useful foliation at large distances even for
$K_{ij} \ne 0$~\cite{Schnetter03a}.

A simple algorithm for pretracking can be described as follows: Define
a parametrised family of CE surfaces $h_p(\theta,\phi)$ which envelop
the source, via
\begin{eqnarray}
  h_p(\theta,\phi) & : & \Theta_{(\ell)}[h_p] = C_p,
\end{eqnarray}
where $p$ labels each member.  Here we introduce the notation $[h]$ to
indicate that a quantity is a functional of the surface function
$h(\theta,\phi)$.  The surface with the smallest positive expansion
$C_p$ forms the ``best guess'' for the location of the AH.  For time
evolutions of vacuum spacetimes, this surface is expected to change
continuously,
and can be followed until the expansion of the surface
reaches zero, at which point it represents the earliest common AH.

\begin{figure}
  \includegraphics[width=0.49\textwidth]{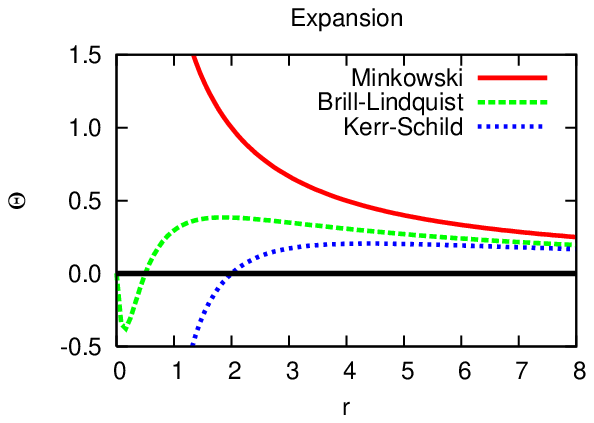}
  \includegraphics[width=0.49\textwidth]{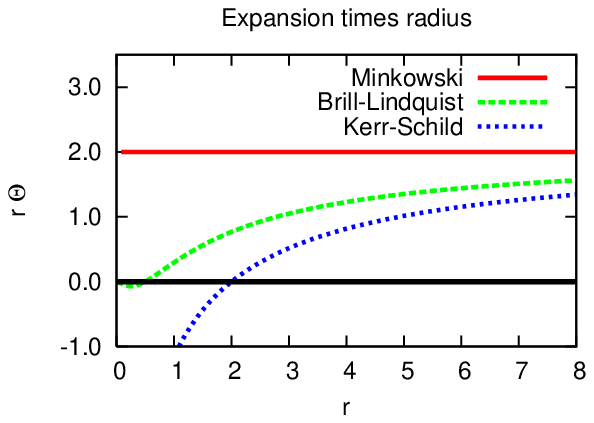}
  \caption{
    Expansion $\Theta$ and the quantity $r \Theta$ vs.\ coordinate
    radius $r$ for flat space in Minkowski coordinates and for a black
    hole in Brill-Lindquist and Kerr-Schild coordinates.  Both
    $\Theta=0$ and $r \Theta=0$ define the location of the apparent
    horizon, which is at different radii in different coordinate
    systems.  While $\Theta$ reaches a maximum at a finite radius, $r
    \Theta$ increases monotonically in spherical symmetry.}
  \label{fig:Theta}
\end{figure}

A difficulty arises from the above definition of the pretracking
surface $h_p$ which introduces some practical complications. While the
expansion is zero at the location of the AH, it also decreases to zero
asymptotically towards spacelike infinity, while being positive in
between.  This is illustrated in Fig.~\ref{fig:Theta} for flat space
in Minkowski coordinates and for a black hole in Brill-Lindquist and
Kerr-Schild coordinates.  For horizon
finding we are more interested in the surfaces most closely enveloping
the source. Thus, the initial guess must be chosen inside the maximum of
$\Theta$ if the method is to avoid converging to one of the asymptotic
surfaces.  To remove this ambiguity, a more suitable function
would increase monotonically outside the common AH. In practice, one
therefore has to use a modified minimisation criterion, as described
below.

\section{Horizon Pretracking}
\label{sec:pretracking}

\subsection{Method}
\label{sec:method}

The pretracking method attempts to find a surface corresponding to
a ``generalised apparent horizon''.
That is, we would like to find a horizon-like surface which envelops
a compact source, even at times when an actual common horizon does not
exist, but which reduces to the apparent horizon when it is present.
The generalised surface is found among a specified parametrised family,
$\{h_p\}$. The pretracking method corresponds to finding
the ``smallest'' member of the family, i.e., the value of the parameter
$p$ such that the surface $h_p(\theta,\phi)$ satisfies
\begin{eqnarray}
  \label{eqn:goal}
  H[h_p] & = & \mathrm{const},
  \\
  G[h_p] & \stackrel{!}{=} & \mathrm{min},
\end{eqnarray}
where the \emph{shape function} $H[h]$ is a generalisation of the
expansion $\Theta_{(\ell)}$, and the \emph{goal function} $G[h]$
specifies what we mean by ``smallest''.  Note that $H$ maps to a
function, while $G$ maps to a scalar. 

The parametrised family $\{h_p\}$ should be defined in such a way as
to contain the common AH if it exists in the given slice.  The shape
function, $H$, specifies the surfaces that we want to find.  We allow
here for a more generic set of surfaces than only CE surfaces.  The
goal function, $G$, defines a suitable notion of ``closeness'' among
the surfaces $H_p$ to an actual apparent horizon. As such, it is
useful to define it in such a way that it evaluates to zero for a
surface which is a common AH, negative within the AH, and
monotonically increasing outside the AH.

We generalise the concept of CE surfaces for a number of pragmatic
reasons. For instance, very distorted surfaces are difficult or
impossible to represent accurately in terms of a function
$h(\theta,\phi)$, so we have found more rounded surfaces work
consistently better. We have experimented with the following shape
functions:
\begin{eqnarray}
  \label{eqn:H-1}
  H_1[h] & := & \Theta_{(\ell)}[h],
  \\
  \label{eqn:H-r}
  H_r[h] & := & \Theta_{(\ell)}[h] \cdot h,
  \\
  \label{eqn:H-r2}
  H_{r^2}[h] & := & \Theta_{(\ell)}[h] \cdot h^2.
\end{eqnarray}
As above, $h(\theta,\phi)$ is the shape of the surface, i.e., the
surface's coordinate radius at the coordinate angles $\theta, \phi$.
The first definition, $H_1$, is simply the surface's expansion, so
that the condition $H_1[h] = \mathrm{const}$ specifies CE surfaces.
We find empirically that the shape functions $H_r$ and $H_{r^2}$ lead
to surfaces with a less distorted coordinate shape.  (This can be seen
e.g.\ in the lower left hand graph of
Figure~\ref{fig:pi0-pretrack-surfaces}.)
The shapes $H_r$ and $H_{r^2}$ are in general not CE
surfaces, but they tend to CE surfaces for large radii and close to
the apparent horizon.

Note that $H_r$ and $H_{r^2}$ are not defined in a covariant manner,
because their definition depends on the coordinate shape.  While we
would prefer covariant shapes, we find that the listed non-covariant shapes
work much better in practice and are computationally more efficient.
Empirically, especially the shape
function $H_r$ leads to a reliable pretracking method.  All of the
above surface shapes are apparent horizons if and only if the shape
function is zero everywhere.

The goal function $G[h]$ must have a minimum for the surface $h_p$
that we want to call ``closest'' to being an AH.  That is, the goal
function has a minimum which is located at the AH when one
exists. We considered the following goal functions:
\begin{eqnarray}
  \label{eqn:G-H}
  G_H[h] & := & \overline{H[h]},
  \\
  \label{eqn:G-Hr}
  G_{Hr}[h] & := & \overline{H[h]} \cdot \overline{h},
  \\
  \label{eqn:G-r}
  G_r[h] & := & \overline{h},
\end{eqnarray}
where the over-bar $\overline{\cdot}$ denotes the average over the
surface.  We take this average as an average over all grid points.
One could alternatively define a covariant average that takes the
two-metric into account, but we did not consider the advantage of this
to be worth the additional complexity in the equations for the
Jacobians (cf.\ Section~\ref{sec:surface-finding} and Appendix
\ref{sec:jacobians}).
For a solution $h$, we require not only a specific value of $G$, but
also that $H[h]$ be constant over the surface, so that e.g.\
$G_{Hr}=0$ implies $H=0$ everywhere.

The combination of the shape function $H_1$ and the goal function
$G_H$ leads to the simple algorithm mentioned in the previous section.
As already discussed, the difficulty with this choice is that even in
spherical symmetry, the goal function is not monotonic, but rather
reaches a maximum at a finite radius.  Thus, this method is not
expected to converge unless an appropriate initial surface, within the
maximum, is chosen. One way to resolve this ambiguity is to redefine
the shape function, for instance via~(\ref{eqn:H-r})
or~(\ref{eqn:H-r2}). For either of these surface functions, the goal
function $G_H$ increases monotonically (at least in spherical
symmetry) and is zero on the AH. The lower graph of
Fig.~\ref{fig:Theta} shows the goal function $G_H$ for the shape
function $H_r$, i.e., the quantity $r \Theta$.

The combination of the shape function $H_1$ and the goal function
$G_{Hr}$ also ensures a monotonically increasing goal function in
spherical symmetry, with the additional advantage that all pretracking
surfaces are CE surfaces, and thus covariantly defined. Unfortunately
we find empirically that this combination does not work as well in
practice as the combination of $H_r$ and $G_H$ (see
Section~\ref{sec:results}).

The goal function $G_r$ measures the average coordinate
radius of a given surface, so that the method selects the
surface with smallest average coordinate radius satisfying
$H[h]=0$. Note however, that the AH does not usually form at the
innermost surface defined by $H$ (as shown in in
Section~\ref{sec:results}).  Thus this goal function cannot reliably
be used for pretracking, although it can still be very useful to study
the behaviour of the innermost CE surface.

Finally, we note that the goal functions $G_{Hr}$ and $G_r$ are not
covariantly defined, as they depend on the coordinate radius of the
surface.  This is not an issue, as the value of the goal function
itself has no relevance other than the value $G[h]=0$, which by
definition is the covariantly defined AH.

\subsection{Surface Finding}
\label{sec:surface-finding}

Finding a surface $h(\theta,\phi)$ that satisfies one of the above
conditions $H[h]=\mathrm{const}$ under a constraint $G[h]=p$ for a
given value $p$ is comparable to finding an apparent
horizon. Efficient implementations of apparent horizon finders have
been developed in recent years~\cite{Schnetter03a,
Thornburg2003:AH-finding}, and it is a straightforward extension to allow
them to look for alternate similarly defined surfaces, such as those
defined by the $H_*$ listed above.  Current efficient AH finders
interpret the surface defining equation as a nonlinear elliptic
equation; they use a Newton-Raphson iteration method to linearise and
then a Krylov subspace method to solve it.

The overall elliptic equation that defines the pretracking surface is
\begin{eqnarray}
  \label{eqn:pretracking}
  H[h](\theta,\phi) - \overline{H[h]} + G[h] - p & = & 0,
\end{eqnarray}
which is fulfilled if and only if $H[h]$ is constant over the surface
and if $G[h]=p$.  
This is because the first term $H[h]$ is the only term that
varies over the surface, and therefore it has to be constant for the
equation to be
fulfilled.  In this case, the expression $H[h] - \overline{H[h]}$
vanishes, which then implies $G[h]=p$.  Discretised, this equation becomes
\begin{eqnarray}
  \label{eqn:equation-simple}
  H_i(h_j) - \frac{1}{N} \sum_k H_k(h_j) + G(h_j) - p & = & 0,
\end{eqnarray}
where $i,j,k$ label grid points, and $N$ is the total number of grid
points on the horizon surface.  The Jacobian of this equation is
\begin{eqnarray}
  \label{eqn:jacobian-simple}
  J_{ij} - \frac{1}{N} \sum_k J_{kj} + \partial_j G,
\end{eqnarray}
where $J_{ij} = \partial H_i / \partial h_j$ is the Jacobian of the
surface shape function $H_i$, and $\partial_j G = \partial G /
\partial h_j$.  For example, when $H$ is the shape function $H_1$, the
Jacobian $J_{ij}$ is the Jacobian
of the original apparent horizon equation.

There is one crucial problem that appears when one attempts to solve
equation~(\ref{eqn:equation-simple}).  The Newton-Raphson method
requires the Jacobian of the equation to be sparsely populated,
otherwise the method will be prohibitively expensive.\footnote{A
  non-local discretisation of the horizon shape, e.g.\ an expansion in
  spherical harmonics, would not require a sparse Jacobian.}
Unfortunately, the Jacobian~(\ref{eqn:jacobian-simple}) is densely
populated due to the term $\sum_k J_{kj}$, which is nonzero for all
values of the indices $i$ and $j$.  For $N$ grid points, this
increases the storage requirements by a factor $O(N)$, and turns the
$O(N^{3/2})$ runtime cost of the solver into a cost of $O(N^3)$.
This is in general not acceptable.

We arrive at a sparse matrix by extending the vector
space of the solution $h_i$ by one element.  Instead
of~(\ref{eqn:equation-simple}), we write the equivalent system
\begin{eqnarray}
  \label{eqn:equation-good}
  H_i(h_j) - C & = & 0,
  \\\nonumber
  C - \frac{1}{N} \sum_k H_k(h_j) + G(h_j) - p & = & 0,
\end{eqnarray}
where we have introduced one additional unknown $C$ and one additional
equation that determines it.  The Jacobian of this equation is
\begin{equation}
  \label{eqn:jacobian-good}
  \left(
    \begin{array}{c|c}
      \displaystyle J_{ij}
      &
      \displaystyle -1
      \\\hline
      \displaystyle - \frac{1}{N} \sum_k J_{kj} + \partial_j G
      &
      \displaystyle +1
    \end{array}
  \right) \quad\textrm{.}
\end{equation}
This Jacobian is now sparsely populated.  In addition to the already
sparsely populated $J_{ij}$, it has one additional fully populated row
and column, so that it has $O(N)$ nonzero entries out of $O(N^2)$
total.  The system~(\ref{eqn:equation-good}) can therefore be solved
efficiently.

We list the full expressions for the Jacobians for the different shape
and goal functions in Appendix~\ref{sec:jacobians}.

\subsection{Pretracking Search}
\label{sec:search}

At each time step during an evolution, pretracking consists of
determining the parameter $p$ which selects a member $h$ of the family
of CE surfaces $\{h_p\}$ that minimises the goal function $G[h]$.  A
convenient initial guess for $p$ is either the value from the last
pretracking time or a manually specified guess, for instance
corresponding to a large sphere.  Because the equation defining
the surface is nonlinear, it is also necessary to specify an initial
guess for the surface shape $h$ itself, again either from a previously
determined surface or manual specification.

We have modified an apparent horizon finder so that we can specify the
desired shape function $H$, the goal function $G$, the desired
parameter value $p$, and an initial guess $h_0$ for the surface.  The
result is either a surface $h$ that satisfies $H[h]=\mathrm{const}$
and $G[h]=p$, or a flag indicating that no such surface could be
found.

The pretracking surface generally exists only for a certain parameter range
$p>p^*$, i.e., only for values of $p$ above some critical parameter
$p^*$. During the search, we keep track of an interval
$[p_\mathrm{min}, p_\mathrm{max}]$ that indicates which values of $p$
are known to fail and succeed, respectively.  We start with a trial
value of $p$ that was the result of the last pretracking time, and
depending on whether the corresponding surface can be found or not, we
set either $p_\mathrm{min}$ or $p_\mathrm{max}$.  We then increase or
decrease $p$ in large steps until we find the other end of the
interval.  Finally, we use a binary search within the interval to find
$p^*$ to a given accuracy $\Delta p$, and call the resulting surface
the pretracking surface at this time.

Each step of the above algorithm, i.e., each check of a parameter
value $p$, is computationally equivalent to finding an apparent
horizon with the advantage of a good initial guess provided by the
previous search. This takes usually less than a second per step on
current hardware, independent of the resolution of the underlying
spacetime. Various trade-offs can be made to improve efficiency.  For
instance, pretracking need not be carried out at every time step of an
evolution, but rather at longer intervals.  We usually determine $p^*$
to only a moderate accuracy at each pretracking time. 

In order to find the common apparent horizon as early as possible, we
follow up with a regular apparent horizon search using the pretracking
surface as initial guess, which is a very good initial guess near the
time when the common AH forms.  This finds the common apparent horizon
even when pretracking is not accurate, and also allows us to pretrack
in larger time intervals. The inaccuracy in the horizon guess
provided by pretracking comes mainly from the fact that failure to
find a pretracking surface for a certain parameter value $p$ does
not necessarily indicate that the surface does not exist. Since
we usually compromise some accuracy for efficiency in the
pretracking search, it may be that we did not iterate long
enough or did not start with a good enough initial guess for
this value of $p$.

\subsection{Apparent Horizon Tracking}

When a common apparent horizon first forms, we find that it bifurcates
into an inner and an outer horizon.  This has, for example, been
described by Thornburg~\cite{Thornburg2003:AH-finding} (cf.\ his
Figures 3 and 4).  If the apparent horizon world tube is spacelike,
then the spacelike hypersurface $\Sigma$ can intersect (``weave'') the
world tube in arbitrary ways, and generically, such intersections will
occur in pairs.
This can be seen e.g.\ in \cite[Fig.~3]{Shapiro85}, which outlines the
trapped region in a spherical collapse.  At early times, the trapped
region's boundary is spacelike.  The outer boundary, which corresponds
to the apparent horizon, approaches the event horizon at late times.
The structure of individual, inner, and outer horizons is also visible
in \cite[Fig.~1]{Bishop82} or \cite[Fig.~1]{Cook92}.

We demonstrate this for the
Brill-Lindquist data in the next section.  For black hole evolutions,
one usually chooses gauge conditions that have the effect of making
the inner horizon quickly shrink in coordinate space, while the outer
horizon initially grows and then stays roughly constant in radius.

Of course, we would like to ensure that horizon tracking follows the
outer horizon, as it is the one of physical interest to observers in
the exterior spacetime. However, unless special measures are taken,
the horizon finder will select a branch at random. We observe that
shortly after bifurcation the
outer horizon is generally less distorted in coordinate shape
in a binary black hole merger.  
Before the inner and outer horizons have clearly separated in
coordinate space, the tracker can also jump between the two branches.
We can make the horizon
tracker select the outer horizon by smoothing its initial guess.
Similarly, we can make it select the inner horizon through a more distorted
initial guess.  Additionally, by slightly enlarging or
shrinking the initial guess, the outer or inner horizons are
preferred, respectively. We implement this by modifying the initial
guess, $h_0$, before each horizon search by the prescription
\begin{eqnarray}
  h_0 & \rightarrow & (1-f) \cdot g \cdot h_0 + f \cdot g \cdot
  \overline{h_0},
\end{eqnarray}
where $\overline{\cdot}$ is the average over the surface.  The
smoothing factor, $f$, should be positive for smoothing and negative
for roughening.  We find that $f=0.25$ is a good smoothing choice for
the particular data we examined.  The enlargement factor $g$ must be
positive, and will grow (or shrink) the surface if it is larger (or
smaller) than one. We find that $g=1.1$ is usually a reasonable value
to ensure that an outer horizon is found. On the other hand, $f=-0.2$
and $g=0.8$ selects the inner horizon in our example of a head-on
collision discussed in the next section.

\section{Applications}
\label{sec:results}

We applied the algorithms described in the previous sections to 3D
simulations of (1) an
axi-symmetric head-on collision and (2) coalescing binary black holes
with angular momentum. For the time evolution we used the
conformal-traceless (BSSN) formalism introduced in~\cite{Nakamura87,
Shibata95, Baumgarte99}.  The particular implementation we use,
including gauges ($1+\log$ lapse and $\Gamma$-driver shift) and
boundary conditions, is described in~\cite{Alcubierre02a}.  The binary
black hole initial data are calculated via the puncture
method~\cite{Brandt97b}.  Spins of the surfaces were measured via the
dynamical horizon formalism, as described
in~\cite{Dreyer-etal-2002-isolated-horizons}.  Our time evolution code is
implemented in the Cactus framework~\cite{Cactusweb}.

Pretracking was performed using extensions to Thornburg's
\texttt{AHFinderDirect} horizon
finder~\cite{Thornburg2003:AH-finding}.  A number of different
shape and goal functions were tested. These are listed in
Table~\ref{tab:methods}, along with corresponding labels which
will be referenced in the plots in this section.

\begin{table}
  \begin{tabular}{c|ccc}
    Label            & Shape     & Goal     & tracked \\[-0.5ex]
                     & function  & function & horizon \\\hline
    \mtheta          & $H_1$     & $G_H$    & outer   \\
    \mthetar         & $H_r$     & $G_H$    & outer   \\
    \mthetarr        & $H_{r^2}$ & $G_H$    & outer   \\
    \mthetarbar      & $H_1$     & $G_{Hr}$ & outer   \\
    \msmallestr      & $H_1$     & $G_r$    & outer   \\
    \minnerthetar    & $H_r$     & $G_H$    & inner   \\
  \end{tabular}
  \caption{Algorithms described in this paper are listed
    according to their use of shape functions and goal functions.
    The methods tested in this paper correspond to the various
    shape and goal functions defined in
    Section~\ref{sec:method}. The names listed in the leftmost column
    are used to label curves in the figures in this section.}
  \label{tab:methods}
\end{table}

\subsection{Head-on Collision}
\label{sec:head-on}

As a first test case, we studied a simple axi-symmetric head-on
collision of two black holes with time-symmetric Brill-Lindquist
initial data~\cite{Brill63}, evolved in 3D.  The black holes have a
mass parameter
$m=1$, and are located on the $z$-axis coordinate locations $z=\pm
1$. The total ADM mass of the system is therefore $M=2$.
The resolution for the particular runs performed here is $\Delta
x=1/8$, and the outer boundaries were placed at coordinate
$x_\mathrm{max}=y_\mathrm{max}=4$ and $z_\mathrm{max}=5$.  We chose a
pretracking surface and apparent horizon resolution of $5^\circ$ and a
pretracking accuracy of $\Delta p = 10^{-4}$.

As the singularities are punctures, we evolved them without excision
by ensuring that their locations are staggered between grid
points~\cite{Alcubierre02a}.  For the lapse, $\alpha$, we used a
$1+\log$ slicing condition, starting from $\alpha=1$, and used normal
coordinates, $\beta^i=0$ throughout the evolution.  The simulation was performed with an
explicit octant symmetry. The model is particularly useful as a test
case, as it begins with disconnected apparent horizons, but very soon
(after $t\approx 1$) forms a common horizon, requiring minimal resources
in terms of computation time or memory for the given resolutions and
boundary locations.

\begin{figure*}
  \includegraphics[width=0.49\textwidth]{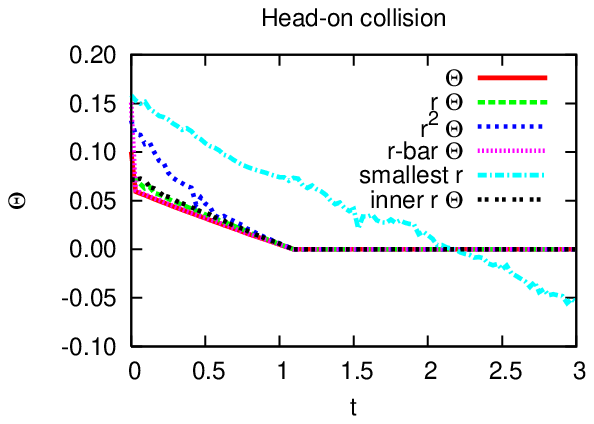}
  \includegraphics[width=0.49\textwidth]{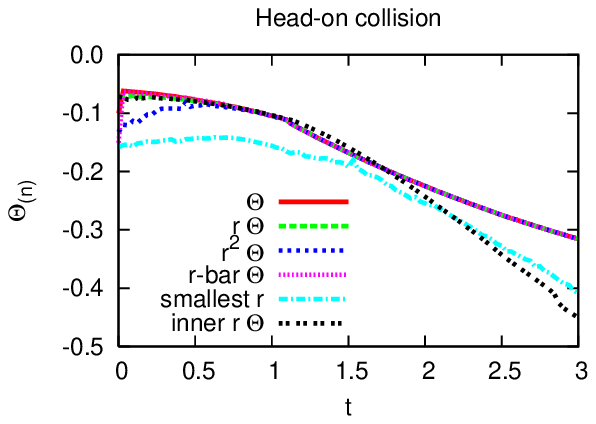}
  \includegraphics[width=0.49\textwidth]{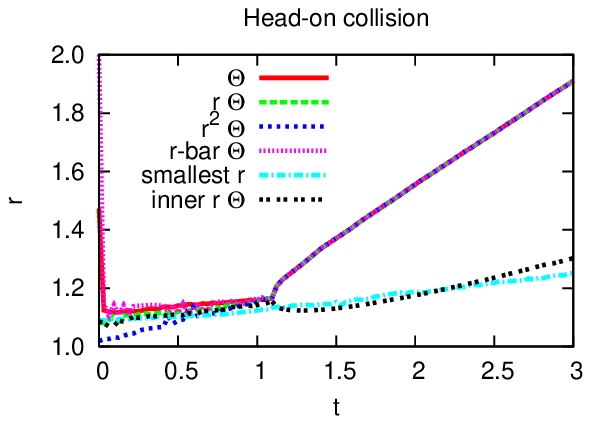}
  \includegraphics[width=0.49\textwidth]{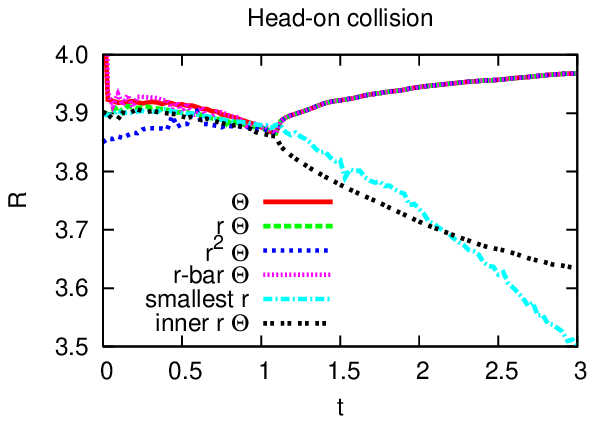}
  \caption{
    Averages of the expansion $\Theta$, inner expansion
    $\Theta_{(n)}$, coordinate radius $r$, and areal radius $R$ for a
    head-on collision vs.\ coordinate time $t$.  Except for the
    \msmallestr\ case, the common apparent horizon is found when
    $\Theta=0$.  Shown are the results for five different pretracking
    algorithms.  The methods labelled \mtheta, \mthetar, \mthetarr,
    and \mthetarbar\ track minimum surfaces of the corresponding
    quantity, as described in the main text.  The \msmallestr\
    algorithm tracks the CE surface with the smallest coordinate
    radius, and fails to converge to the apparent horizon.  Finally,
    \minnerthetar\ uses the same pretracking algorithm as \mthetar,
    but locks onto inner instead of outer horizons.
    The roughly parabolically shaped region between the inner and
    outer horizons in the lower left hand graph contains trapped
    surfaces.}
  \label{fig:HeadOn}
\end{figure*}

Figure~\ref{fig:HeadOn} shows various pretracking results from this
evolution.  The top left graph shows the average of the expansion
$\Theta$ versus time for pretracking surfaces from the various
algorithms listed in Table~\ref{tab:methods}.  The algorithms \mtheta,
\mthetar, \mthetarr, and \mthetarbar\ all find the common apparent
horizon at identical times, around $t=1$.  (However, the algorithm
\mthetarbar\ unfortunately fails in more realistic simulations; see
e.g.\ Figure~\ref{fig:pi0-pretrack-surfaces}.)  The algorithm
\msmallestr\ fails to find the common horizon.  It locates the CE
surface with smallest average radius, a surface which has a positive
average expansion even after a common apparent horizon has formed.
This surface actually intersects the pretracking surfaces defined by
the other algorithms, and seems ``smallest'' due to the averaging of
the radius over the surface which includes a narrow throat.

The top right plot shows the average of the expansion of the ingoing
null normal $\Theta_{(n)}$ which must be negative for apparent
horizons. The lower two graphs show the average coordinate radius
(left) and areal radius (right) versus time of the pretracking
surface. After an initial transient, all of the pretracking algorithms behave
similarly, with the exception of the \msmallestr, which fails
as a pretracking surface. The gradual growth in coordinate radius of
the common apparent horizon is caused by the shift condition; a zero
shift is unsuitable for a long-term evolution since the black hole
eventually encompasses the grid.  Note that the
\minnerthetar\ algorithm tracks as desired the inner apparent
horizon (as described in the previous section), which has a smaller
radius.

\begin{figure}
  \begin{flushleft}
    \includegraphics[width=0.49\textwidth]{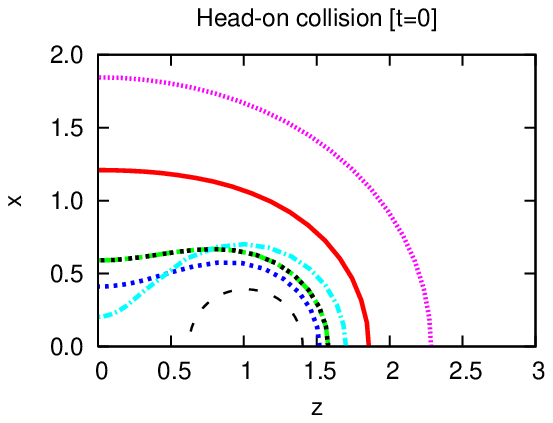}
    \includegraphics[width=0.49\textwidth]{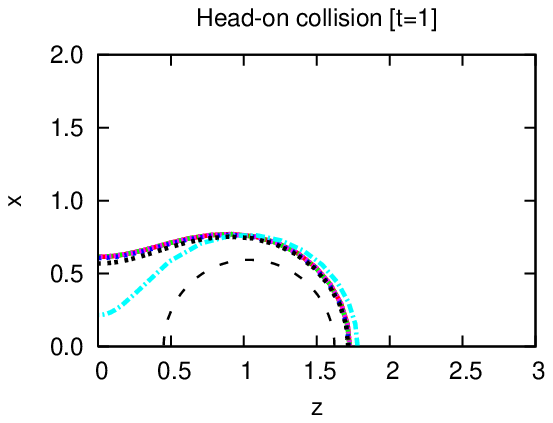}
    \includegraphics[width=0.49\textwidth]{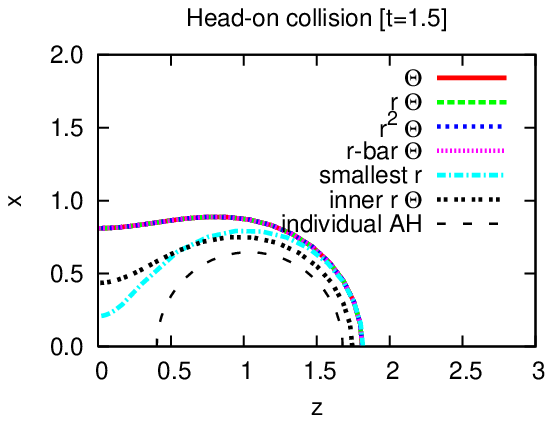}
  \end{flushleft}
  \caption{
    Shapes of the pretracking surfaces for various pretracking methods
    at different times.  The coordinate time $t=0$ shows the initial
    data. The time $t=1$
    is shortly before the common apparent horizon is detected.  Here
    the different pretracking algorithms already produce very similar
    results.  At $t=1.5$ a common apparent horizon has formed, and the
    inner and outer horizons have separated.  The individual apparent
    horizons are also shown for comparison with the pretracking
    surfaces.
    The region between the inner and outer horizons in the lowermost
    graph contains trapped surfaces.}
  \label{fig:HeadOn-shapes}
\end{figure}

Before a common horizon is formed, the different pretracking
algorithms actually determine different
surfaces. Figure~\ref{fig:HeadOn-shapes} shows the shapes of the
pretracking surfaces at $t=0$, $1$, and $1.5$ in one quadrant of the
$xz$-plane.  Although the three successful pretracking algorithms
determine rather different surfaces initially, they have converged
by $t=1$, shortly before the first common apparent horizon is found.  At
$t=1.5$ the first three algorithms continue to track the same outer
horizon surface, while the difference between the outer and inner
apparent horizon (as tracked by the \minnerthetar\ method) is
clearly visible. The algorithm \msmallestr, which fails,
tracks a very different surface.  This
surface is nevertheless interesting because it is, up to numerical
errors, the smallest CE surface that exists within the given slices.

\subsubsection{Evolution of CE Surfaces}

\begin{figure*}
  \begin{flushleft}
    \includegraphics[width=0.49\textwidth]{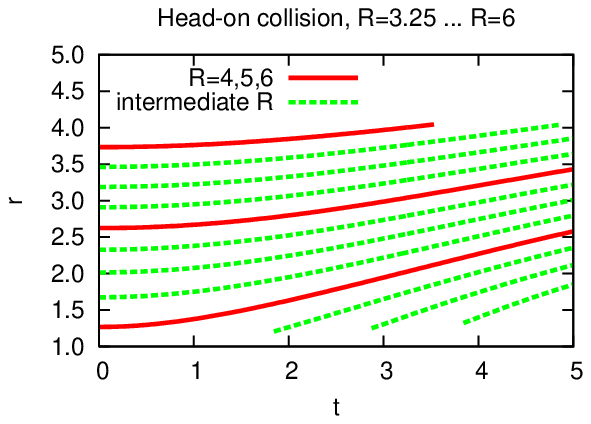}
    \includegraphics[width=0.49\textwidth]{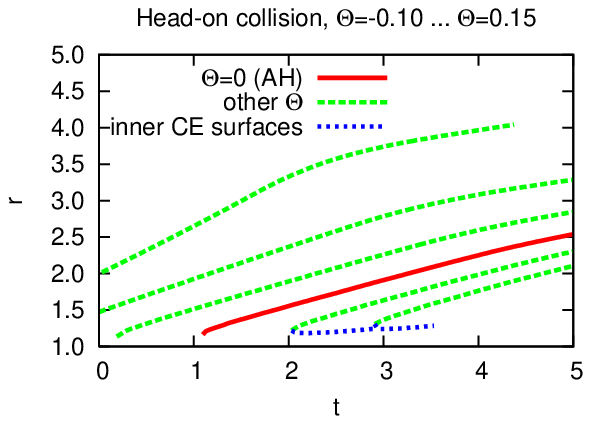}
    \includegraphics[width=0.49\textwidth]{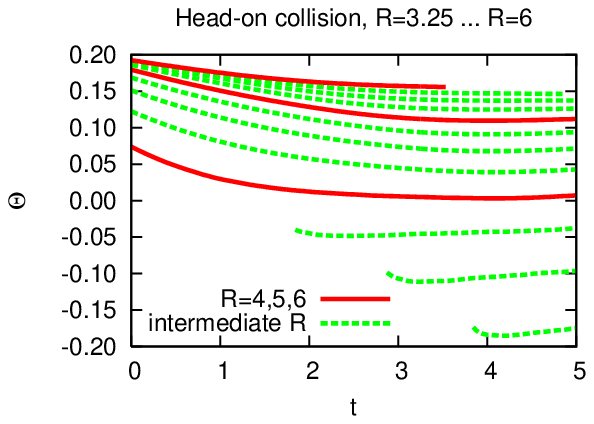}
    \includegraphics[width=0.49\textwidth]{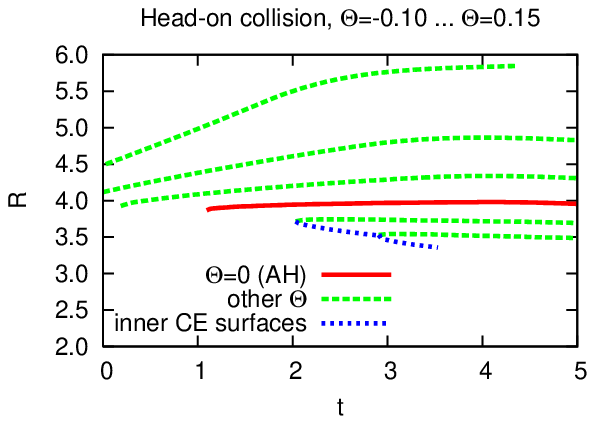}
  \end{flushleft}

  \caption{
  Left (top and bottom): Time evolution of various CE surfaces with
  given areal radii $R$.  The graphs show CE surfaces with areal radii
  from $3.25$ to $6$ (from bottom to top).  The upper left graph shows
  that these surfaces grow with time, the lower left graph shows that
  their expansion decreases with time.  The largest CE surface is not
  found after $t \approx 3.5$ any more because it reached the outer
  boundary.  The three smallest CE surfaces only begin to exist after
  some time.
  \\
  Right (top and bottom): CE surfaces with given expansions $\Theta$.
  The coordinate radii of
  the surfaces increase with time, and new surfaces come into
  existence near the inner boundary where the CE surfaces cease to
  exist.  The common AH is one of these surfaces.  For the smallest
  new CE surfaces also the corresponding inner surfaces (with the same
  expansion) are shown.}
  \label{fig:HeadOn-expansion}
\end{figure*}

It is interesting to study the behaviour of various CE surfaces
independent of the pretracking algorithm.
Figure~\ref{fig:HeadOn-expansion} compares how CE surfaces with constant
areal radius $R$ and CE surfaces with constant expansion $\Theta$
evolve in time. Inside an inner cutoff, the CE surface equation has no
solution. We are using normal coordinates, and constant areal radius
surfaces grow in coordinate space. New common CE surfaces with
smaller areal radii and smaller expansions begin to exist near the
inner cutoff.  One of these new surfaces is the common apparent
horizon with expansion $\Theta=0$.

\subsubsection{Convergence Properties of the Method}

We used the above setup to test convergence of the pretracking code on
the initial data set for the method \mthetar.  There are several
numerical steps involved in pretracking, namely first the time
evolution of the spacetime, then an interpolation to the pretracking
surface, then the solution of an elliptic equation on the pretracking
surface, and finally the pretracking search for the critical parameter
value $p^*$.  Each of these steps may or may not converge to a certain
appropriate order.  The intention here is to test the accuracy of the
pretracking iterations, assuming that the remaining steps are already
convergent, in order to demonstrate that this pretracking search is a
well-defined problem, i.e., that such a critical value $p^*$ does indeed
exist independent of the resolution given appropriately defined shape
and goal functions.

The convergence of the interpolation and the apparent horizon finder
has already been reported in \cite{Thornburg2003:AH-finding}, and our
modifications to make it find pretracking surfaces instead of apparent
horizons do not affect this.  We omit studying a time evolution by
considering only Brill-Lindquist initial data.  We only change the
resolution $\Delta x$ of the data given on the 3D hypersurface, and
calculate the critical parameter $p^*$ as a function of $\Delta x$.
We leave the resolution of the pretracking surface constant at
$2.25^\circ$, and set the pretracking tolerance to $\Delta p =
10^{-8}$.

The resolutions $\Delta x$, resulting critical parameters $p^*$,
and areal radii $R$, of the numerically determined surfaces are given
in Table~\ref{tab:convergence}.  The results converge for higher
resolutions, but a fourth order convergence is not evident.
This is likely caused by the fact that
finding a surface depends not only  on the desired
goal function value $p$, but also on the initial guess $h_0$ for the
surface.  This makes it in practice very difficult to determine the
critical parameter value $p^*$ rigorously.  The given values $p^*$ and
$R$ have therefore errors that are unrelated to the grid spacing,
which makes a convergence test problematic.  We take here a pragmatic
approach and remark that pretracking does in practice lead to a
reliable detection of the common apparent horizon immediately after it
forms.  The precise value of $p^*$ is here
not of interest.

\begin{table}
  \begin{tabular}{c|cc}
    Grid spacing & Parameter & Areal radius \\
    $\Delta x$   & $p^*$     & $R$          \\\hline
    $1/4$        & $0.05498$ & $3.880$      \\
    $1/5$        & $0.05665$ & $3.893$      \\
    $1/8$        & $0.05540$ & $3.887$      \\
    $1/10$       & $0.05483$ & $3.889$      \\
    $1/16$       & $0.05507$ & $3.885$      \\
    $1/20$       & $0.05508$ & $3.882$
  \end{tabular}
  \caption{
    Convergence test for different spatial resolutions.  The values
    seem to approach a certain limit when the resolution is increased,
    indicating that the limit $\lim_{\Delta x \to 0} p^*$ does exist.}
  \label{tab:convergence}
\end{table}

\subsubsection{Pretracking Efficiency}

\begin{figure}
  \includegraphics[width=0.49\textwidth]{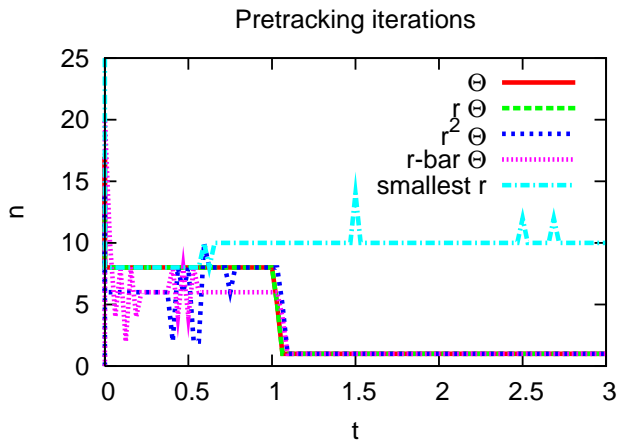}
  \caption{
    Number of pretracking iterations, which has the same cost as the same
    number of AH searches, for different pretracking methods at
    different times.  The pretracking accuracy was $\Delta p =
    10^{-4}$; this means that the value $p^*$ of the pretracking
    surfaces at each time is only accurate up to this error (see
    section~\ref{sec:search}).}
  \label{fig:HeadOn-iters}
\end{figure}

We show in Fig.~\ref{fig:HeadOn-iters} the cost of pretracking.  As
described in Section~\ref{sec:search}, pretracking iterates over
several surfaces until a surface with a certain minimum property has
been found.  Each such pretracking iteration is about as expensive as
finding an apparent horizon.
A ``typical'' simulation of our group runs in parallel on about 30
processors and needs about 20 seconds for each evolution time step.
Finding an apparent horizon with a good initial guess takes about one
second, i.e., about 5\% of the evolution step.
The time spent in the apparent horizon finder is
approximately independent of the resolution of the 3D simulation
and the number of processors.
The efficiency of our fast horizon finder is described in detail
in~\cite{Thornburg2003:AH-finding}.

In this simulation, the successful pretracking
methods take about eight iterations during pretracking, and one
iteration while tracking the horizon later.  We find that it is in
general sufficient to pretrack only every $10$ iterations or fewer, and
the cost of pretracking is therefore acceptable.

\subsection{Coalescing Black Holes with Angular Momentum}

As a further application for pretracking we consider an inspiralling
black hole setup.  We emphasise that pretracking is of high importance
in these cases, where the two black holes can stay apart for
times of more than $100\,M$ (see e.g.~\cite{Bruegmann:2003aw}).  Pretracking
provides valuable information about impending apparent horizon
formation as well as the necessary initial guess for the apparent
horizon finder, or it can (with some confidence) exclude that a common
AH exists.

The model we consider is the innermost stable circular orbit as
predicted in \cite{Baumgarte00a}, which applies the effective
potential techniques of
\cite{Cook94} to puncture initial data \cite{Brandt97b}. Previous
simulations with this dataset suggest that they perform around
a half orbit before a common horizon
first appears~\cite{Baker:2002qf, Alcubierre2003:pre-ISCO-coalescence-times}.
The initial data contain two punctures with
a proper horizon separation of $L=4.99\,M$, angular momentum $J=0.78\,M^2$ and
angular velocity $\Omega=0.17/M$, where $M$ is the ADM mass of the
system.  This dataset was studied in detail
in \cite{Alcubierre2003:pre-ISCO-coalescence-times}.  This model is
the first in the sequence (QC-0) studied with the Lazarus perturbative
matching technique in \cite{Baker:2002qf}.
The QC-0 model is attractive as a test-bed for pretracking since a
common apparent horizon is found at around $t=16.44\,M$, while the
simulation continues accurately beyond $t=40\,M$.

In this model, an initial guess for the common apparent horizon with a
spherical form will find the common AH at the same iteration as
pretracking if the chosen radius is accurate to about 10\%.  For black
hole simulations that start out with a larger separation, the initial
guess usually has to be ellipsoidal in order to find the common AH at
all.  In that case, all axis lengths have to be guessed correctly to
the above accuray.

The evolutions were performed in bitant symmetry, i.e., with a
reflection symmetry about $z=0$ plane.  We used a grid containing $256\times
256\times 128$ points and a resolution of $\Delta x=0.06\,M$, which
places the outer boundary at only about $7.68\,M$ for this test run.  The
evolutions were stopped at $t=22\,M$ since we were not interested in
further tracking the evolution of the common apparent horizon and its
ring-down.  During the evolution we used a co-rotation shift to keep the
black holes from moving across the grid
\cite{Bruegmann:2003aw,Alcubierre2003:co-rotating-shift}, which is
essential for long-term stable evolutions.  We used a $1+\log$ slicing
and a $\Gamma$-driver shift condition \cite{Alcubierre02a}.

\begin{figure*}
  \includegraphics[width=0.65\textwidth]{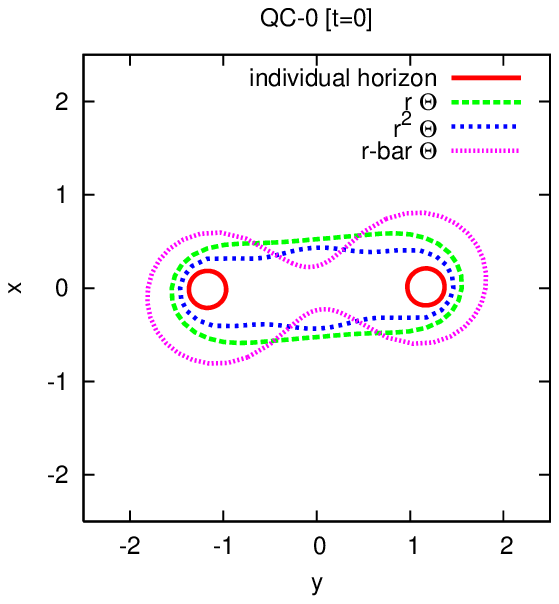}\hspace*{-0.2\textwidth}
  \includegraphics[width=0.65\textwidth]{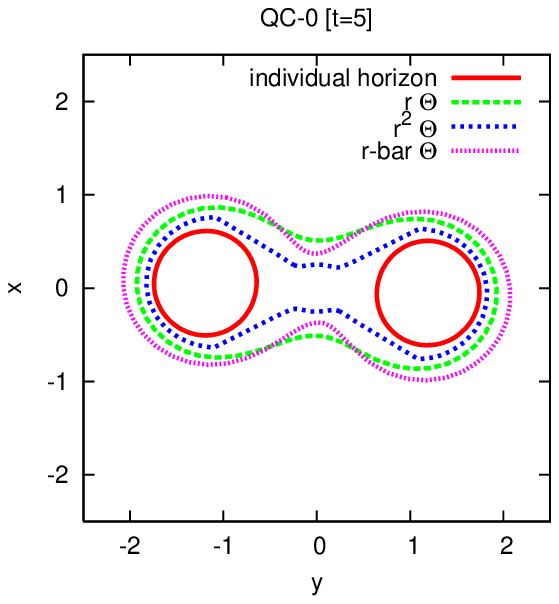}\hspace*{-0.2\textwidth}
  \\
  \includegraphics[width=0.65\textwidth]{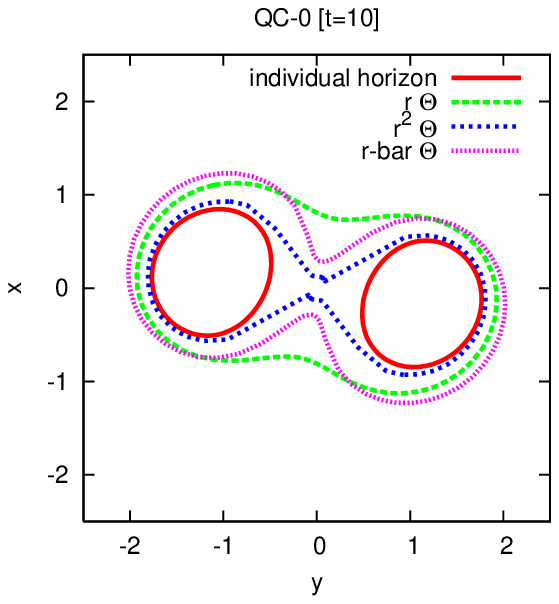}\hspace*{-0.2\textwidth}
  \includegraphics[width=0.65\textwidth]{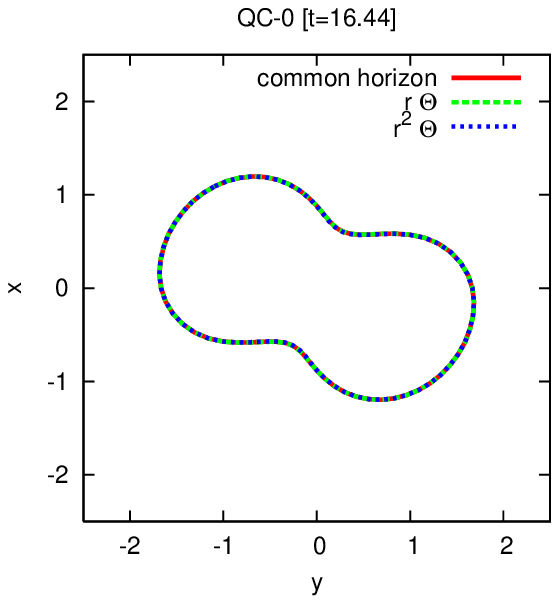}\hspace*{-0.2\textwidth}
  \caption{
    Time evolution of the different pretracking surfaces in the
    $xy$-plane for the inspiralling QC-0 model.  Shown are
    individual and common apparent horizons and
    the pretracking surfaces found using different pretracking
    algorithms.
    A common horizon first appears at $t=16.44M$, shown in the lower
    right plot. It does not show the \mthetarbar\ surface because
    it could not be found at that time.}
  \label{fig:pi0-pretrack-surfaces}
\end{figure*}

Figure~\ref{fig:pi0-pretrack-surfaces} shows the evolution of the
pretracking surfaces found by different algorithms.  The upper left plot shows
the various pretracking surfaces at
the initial time.\footnote{The simplest
method, \mtheta, did not find an appropriate pretracking surface in
this case, and is not shown.}
It is obvious that the different algorithms are
locating very different surfaces.  Note that the
covariant method \mthetarbar\ has the most distorted shape.  The
bump in the waist of the \mthetarr\ surface at $t=5\,M$ and $t=10\,M$ is
caused by insufficient resolution in the surface; our explicit
$h(\theta,\phi)$ representation is not well adapted to a peanut-shaped
surface. The surface evolution can still be tracked, however, and
later approaches the correct AH shape.

After $t=12.75\,M$, the method \mthetarbar\ fails to locate a
surface.  It succeeds again at $t=16.79\,M$, but it has then locked
onto a surface that is inside the AH.  We assume that this is due to
the larger coordinate distortion of this surface, which has a much
more pronounced waist than the other methods.
The individual horizons are
lost after $t=14.34\,M$.  At $t=16.44\,M$ a common apparent horizon is
found for the first time.  The lower right hand graph compares
different surfaces at that time and demonstrates that the succeeding
pretracking methods have converged to the same surface.

\begin{figure}
  \includegraphics[width=0.49\textwidth]{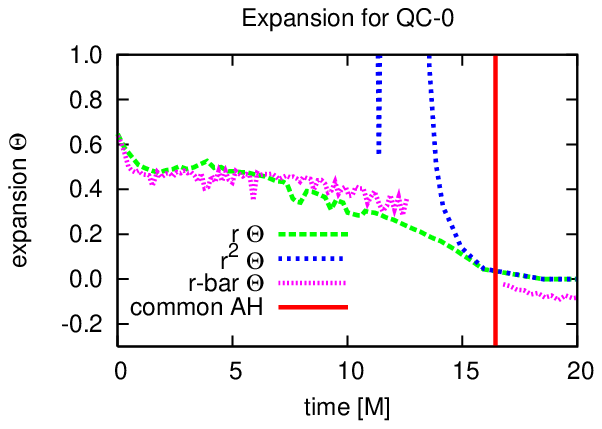}
  \includegraphics[width=0.49\textwidth]{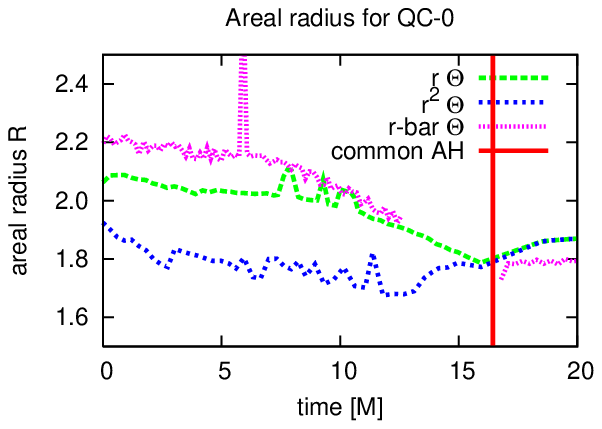}
  \caption{
    Average expansion $\Theta$ and areal radius $R$ for the
    inspiralling black hole evolution.  The average expansion for
    \mthetarr\ is larger than can be shown on this
    scale.  This is due to numerical errors, as explained in the main
    text.  The method \mthetarbar\ fails after $t=12.75\,M$; it
    succeeds again after the common AH has formed, but tracks a
    surface that is not the AH.}
  \label{fig:pi0-pretrack-measures}
\end{figure}

Figure~\ref{fig:pi0-pretrack-measures} shows the average expansions
and the areal radii of the pretracking surfaces.  The time at which
the common AH is found is marked with a vertical line.  The different
pretracking methods track different surfaces during most of the
evolution, but these surfaces converge about $1.5\,M$
before the common AH is found.  The pretracking surfaces approach the
common apparent horizon smoothly, and they accurately predict both its shape and
its area.  Pretracking gives here an early indication of
the time at which the common AH forms.
After the common AH has formed, the pretracking methods \mthetar\ and
\mthetarr\ lock onto the AH.  The method \mthetarbar, which
failed to locate a surface for some time, afterwards tracks a different
surface.  Both methods \mthetar\ and \mthetarr\ are thus reliable
pretracking methods for this evolution.

\begin{figure}
  \includegraphics[width=0.49\textwidth]{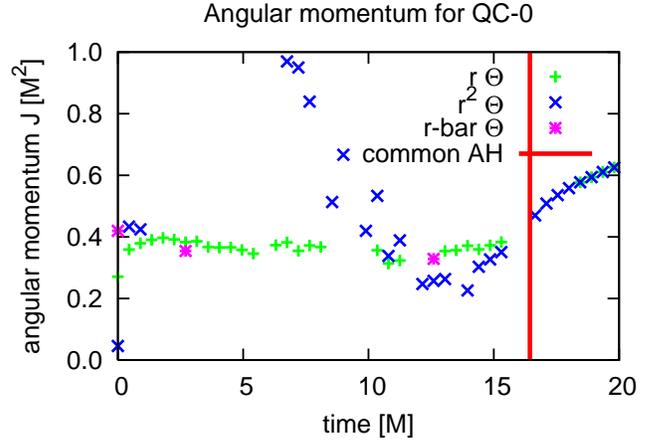}
  \caption{
    Angular momentum $J$ of the pretracking surfaces versus time,
    measured with the dynamical horizon framework.  The
    angular momentum of the common AH increases directly after it has
    formed, indicating that it accretes both energy and angular
    momentum initially.  In the few cases where a value for $J$ could
    be found for the \mthetarbar\ pretracking surfaces, they agree
    well with the \mthetar\ method.  In some cases, no angular
    momentum could be measured because no approximate Killing vector
    could be found on the pretracking surface at that time.}
  \label{fig:pi0-pretrack-spins}
\end{figure}

We measured the angular momentum of the pretracking surfaces using the
dynamical horizon framework~\cite{Ashtekar:2004cn}; our implementation
is described in~\cite{Dreyer-etal-2002-isolated-horizons}.  We show
the results in Fig.~\ref{fig:pi0-pretrack-spins}.
Note that this angular momentum
measure is a quasi-local quantity and therefore is defined on any
surface, not only apparent horizons.  However, our current implementation is
restricted to cases that are close to having an axial Killing vector.
In dynamical spacetimes this condition
will only be satisfied approximately.  Yet in the case of
this merger simulation both successful pretracking methods
\mthetar\ and \mthetarr\ predict the angular momentum of the common
apparent horizon.

We conclude that pretracking works as an effective analysis tool
for this binary black hole
collision.  The time of merger is predicted by the rate in which the
expansion of the pretracking surfaces approaches zero, and the
pretracking surfaces give a good estimate for the shape, area, and
angular momentum for the future common apparent horizon.

\section{Conclusion}

Horizon pretracking is an effective method for predicting the location
and shape of an emerging common horizon, as well as yielding important
physical information during the course of a spacetime evolution. The
simple cases tested here indicate that it is applicable to a
variety of scenarios, such as black hole mergers, where the formation
of a horizon is expected, but its location and shape are not known
\emph{a priori}.

Various practical issues need to be taken into consideration when
choosing an appropriate surface family for pretracking. A family of CE
surfaces parametrised by their expansion suffers some drawbacks, since
the expansion does not increase monotonically. However, a simple
modification such as the multiplication by a coordinate radius leads
to an effective practical algorithm. We have shown that several
methods for defining the pretracking surface all lead to procedures
which accurately determine the common AH at the time of its first
formation.

As the lifetimes of black hole evolutions with dynamic gauges are
extended, this ability to predict the common horizon shape will become
particularly important. Apparent horizon finders require a reasonable
initial guess in order to converge. After a long period of evolution
of separated black holes, the initial common horizon shape is
difficult to predict. As a result, a common AH may not be found, or at
least not at the earliest possible moment. This can lead to the
impression that one is evolving separated black holes when
they have actually already merged.
Valuable physical information, which can be
determined from the shape and oscillations of the common horizon, will
be lost. On a practical level, finding the earliest common AH gives
important information as to the causality of the spacetime and
allows more effective application of certain numerical techniques, such as
singularity excision.

We observe that new trapped surfaces form in pairs of an outer and an
inner surface. A given initial guess for the horizon surface will find
only one of these surfaces, which tend to diverge quickly in
coordinate space. Since initially they are identical, however, an AH
finding algorithm will converge to either the inner or the outer
depending on its initial guess. This is particularly true if the
common AH is found very early (as is the intention of the pretracking
method). We have given a simple algorithm that helps selecting the
outer, generally less distorted surface, which is the astrophysically
interesting apparent horizon. This algorithm
does not depend on pretracking and is useful in general.

We emphasise here that valuable information can be gained from the
study of foliations of CE (and related) surfaces in a slice, beyond
the commonly studied apparent horizons. 
The dynamical horizon formalism defines a quasi-local measure of the
spin that can be evaluated on any closed surface, and the
three-covariantly defined CE surfaces form a natural choice.
Gauge conditions can also be designed to reduce
distortions in such surfaces, or hold them in place on the grid,
potentially simplifying the dynamics.

The pretracking method is efficient enough to be applied as a regular
analysis tool during large simulations. It is already being used in a
variety of simulations of binary puncture and thin sandwich data
evolutions, and is finding further application in hydrodynamical
collapse scenarios.

\begin{acknowledgments}
  We would like to thank Ian Hawke and Jonathan Thornburg for useful
  discussions and for advice on the use of the \texttt{AHFinderDirect}
  code.  The extensions to \texttt{AHFinderDirect} for pretracking
  have been contributed back to the public archive and are available
  for download.  Results for this paper were obtained on AEI's
  \texttt{Peyote} cluster and using computing time allocations at the
  NCSA.  We use Cactus and the \texttt{CactusEinstein} infrastructure
  with a number of locally developed thorns.  ES and FH are funded by
  the DFG's special research centre SFB TR/7 ``gravitational wave
  astronomy''.
\end{acknowledgments}

\appendix

\section{Expressions for the Jacobians}
\label{sec:jacobians}

We list here for the reader's convenience the equations and Jacobians
for the pretracking methods that we use in this paper.  See table
\ref{tab:methods} for the combinations of shape and goal functions and
the names that we use for these combinations.

We assume that a desired goal function value $p$ is given, and a
corresponding surface $h$ is to be determined by the combination of
the shape and the goal function.  For all methods, we list first the
system of equations that defines the surface, and then give the
Jacobian of this system.  The intermediate quantity $C$ is also
determined through these equations, but its value is irrelevant.

We denote the expansion of the surface $h$ with $\Theta_i$, and the
expansion's Jacobian as $J_{ij} = \partial \Theta_i / \partial h_j$.
These expressions are e.g.\ given in \cite{Schnetter03a} or
\cite{Thornburg2003:AH-finding}.  $\delta_{ij}$ is the Kronecker delta
symbol.

\begin{description}

\item[Method \mtheta:] shape function $H_1$, goal function $G_H$:

\begin{eqnarray}
  \Theta_i - C & = & 0
  \\\nonumber
  C - p & = & 0
\end{eqnarray}
\begin{equation}
  \left(
    \begin{array}{c|c}
      \displaystyle J_{ij}
      &
      \displaystyle -1
      \\\hline
      \displaystyle 0
      &
      \displaystyle +1
    \end{array}
  \right)
\end{equation}

\item[Method \mthetar:] shape function $H_r$, goal function $G_H$:
\begin{eqnarray}
  \Theta_i h_i - C & = & 0
  \\\nonumber
  C - p & = & 0
\end{eqnarray}
\begin{equation}
  \left(
    \begin{array}{c|c}
      \displaystyle J_{ij} h_i + \Theta_i \delta_{ij}
      &
      \displaystyle -1
      \\\hline
      \displaystyle 0
      &
      \displaystyle +1
    \end{array}
  \right)
\end{equation}

\item[Method \mthetarr:] shape function $H_{r^2}$, goal function $G_H$:
\begin{eqnarray}
  \Theta_i h_i^2 - C & = & 0
  \\\nonumber
  C - p & = & 0
\end{eqnarray}
\begin{equation}
  \left(
    \begin{array}{c|c}
      \displaystyle J_{ij} h_i^2 + 2 \Theta_i h_i \delta_{ij}
      &
      \displaystyle -1
      \\\hline
      \displaystyle 0
      &
      \displaystyle +1
    \end{array}
  \right)
\end{equation}

\item[Method \mthetarbar:] shape function $H_1$, goal function $G_{Hr}$:
\begin{eqnarray}
  \Theta_i - C & = & 0
  \\\nonumber
  C - \frac{1}{N} \sum_k \Theta_k + \frac{1}{N} \sum_k
  \Theta_k \cdot \frac{1}{N} \sum_k h_k - p & = & 0
\end{eqnarray}
\begin{equation}
  \left(
    \begin{array}{c|c}
      \displaystyle J_{ij}
      &
      \displaystyle -1
      \\\hline
      \displaystyle - \frac{1}{N} \sum_k J_{kj} + \frac{1}{N} \sum_k
      J_{kj} \cdot \frac{1}{N} \sum_k h_k + \frac{1}{N} \sum_k
      \Theta_k \cdot \frac{1}{N}
      &
      \displaystyle +1
    \end{array}
  \right)
\end{equation}

\item[Method \msmallestr:] shape function $H_1$, goal function $G_r$:
\begin{eqnarray}
  \Theta_i - C & = & 0
  \\\nonumber
  C - \frac{1}{N} \sum_k \Theta_k + \frac{1}{N} \sum_k h_k - p &
  = & 0
\end{eqnarray}
\begin{equation}
  \left(
    \begin{array}{c|c}
      \displaystyle J_{ij}
      &
      \displaystyle -1
      \\\hline
      \displaystyle - \frac{1}{N} \sum_k J_{kj} + \frac{1}{N}
      &
      \displaystyle +1
    \end{array}
  \right)
\end{equation}

\item[Method \minnerthetar:] same as \mthetar\ above

\end{description}

\bibliography{bibtex/references}

\end{document}